\documentclass[a4paper]{jpconf}
\usepackage{graphicx}
\begin{document}
\title{Ferromagnetic transition in the double-exchange model on the pyrochlore lattice}

\author{Yukitoshi Motome$^1$ and Nobuo Furukawa$^{2,3}$}

\address{$^1$Department of Applied Physics, University of Tokyo, Tokyo, Japan}
\address{$^2$Department of Physics and Mathematics, Aoyama Gakuin University, Kanagawa, Japan}
\address{$^3$Multiferroics Project, ERATO, Japan Science and Technology Agency (JST)}

\ead{motome@ap.t.u-tokyo.ac.jp}




\begin{abstract}
The double-exchange model, which has been extensively studied in the context of
colossal magneto-resistance in perovskite manganese oxides, 
is known to exhibit a ferromagnetic metallic state at low temperatures
because of the interplay between localized moments and itinerant electrons 
through the Hund's-rule coupling.
Here we investigate numerically the ferromagnetic transition
in the double-exchange model defined on the frustrated pyrochlore lattice 
as a simple model for ferromagnetic pyrochlore oxides.
We demonstrate that the finite-size corrections are largely reduced 
by implementing averages over the twisted boundary conditions 
in the Monte Carlo simulation,
which enables to estimate the ferromagnetic transition temperature
in relatively small size clusters.
The estimate is compared with that for the non-frustrated cubic lattice system.
\end{abstract}

\section{Introduction}
The double-exchange (DE) model is a minimal model 
which explicitly incorporates the interplay 
between itinerant electrons and localized magnetic moments.
The Hamiltonian is given by
\begin{equation}
{\cal H} = - \sum_{\langle ij \rangle \sigma}
t \, ( c_{i \sigma}^\dagger c_{j \sigma} + {\rm h.c.} )
- J_{\rm H} \sum_i {\mathbf s}_i \cdot {\mathbf S}_i - \mu \sum_i n_i,
\label{eq:H}
\end{equation}
where $c_{i \sigma}$ ($c_{i \sigma}^\dagger$) is an annihilation (creation) operator
of an electron with spin $\sigma$ at site $i$, 
$t$ is the transfer integral for nearest-neighbor sites $\langle ij \rangle$,
${\mathbf s}_i$ and ${\mathbf S}_i$ are spin operators of itinerant electron
and localized spin, respectively, which are coupled by the Hund's-rule coupling $J_{\rm H}$,
$n_i = \sum_{\sigma} c_{i \sigma}^\dagger c_{i \sigma}$ is the density operator, 
and $\mu$ is the chemical potential. 
The localized spins ${\mathbf S}_i$ are treated as classical vectors. 
The model was originally introduced by Zener
\cite{Zener1951}, 
and has been studied for understanding the physics of perovskite manganese oxides
\cite{Anderson1955,deGennes1960}.
In particular, the rediscovery of colossal magneto-resistance phenomena have stimulated
extensive studies for the DE systems,
including some extensions of the model 
such as the super-exchange interaction between localized moments, 
orbital degeneracy of itinerant electrons, and electron-phonon couplings
\cite{Kaplan1999,Tokura2000}.

The model (\ref{eq:H}) is known to exhibit a ferromagnetic metallic state at low temperature ($T$)
to gain the kinetic energy of electrons by aligning localized moments in parallel.
This is called the DE mechanism
\cite{Zener1951}. 
There have been many efforts to estimate the ferromagnetic transition temperature $T_{\rm c}$
\cite{Motome2003}.
For the model 
on the three-dimensional cubic lattice, 
$T_{\rm c}$ was determined precisely by a large-scale Monte Carlo simulation: 
For example, $T_{\rm c}/t = 0.136(2)$ at the electron density $n=0.5$
in the limit of $J_{\rm H}/t \to \infty$
\cite{Motome2003b}.

\begin{figure}[h]
\begin{minipage}{37pc}
\includegraphics[width=24pc]{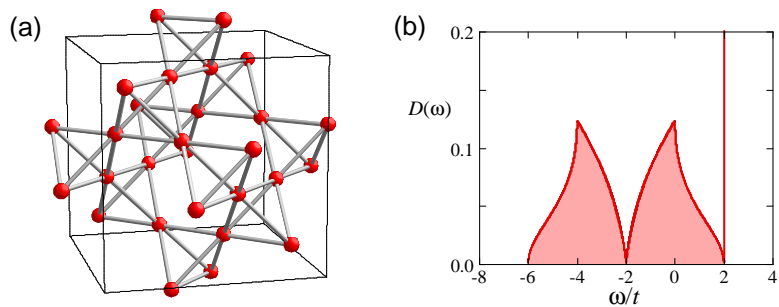}
\caption{
(a) Pyrochlore lattice structure. The box indicates the cubic unit cell. 
(b) Density of states per site for the non-interacting model
on the pyrochlore lattice.
}
\label{fig1}
\end{minipage}
\end{figure}

In this contribution, we present our numerical results for the ferromagnetic transition 
when the model is defined on the geometrically frustrated lattice structure.
Among many frustrated lattice structures, we consider the pyrochlore lattice, 
which is a three-dimensional network of corner-sharing tetrahedra
as shown in Fig.~\ref{fig1}(a).
One of the experimental motivations is found in 
a family of Mo pyrochlore oxides {\it R}$_2$Mo$_{2}$O$_{7}$:
When the ionic size of rare earth element {\it R} is relatively large such as {\it R} = Nd and Sm, 
the compounds become ferromagnetic metal at low $T$
\cite{Greedan1987,Katsufuji2000}, 
and it was pointed out by the first principle calculations 
that the DE mechanism plays a key role for this behavior 
\cite{Solovyev2003}. 
As a first step toward the understanding of thermodynamic properties of 
the pyrochlore systems, 
below we will investigate the ferromagnetic transition by Monte Carlo calculations 
for the simplest case, i.e., the model (\ref{eq:H}) 
in the limit of $J_{\rm H}/t \to \infty$.

\section{Monte Carlo Simulation with Averaging over Twisted Boundary Conditions}
We employ a Monte Carlo method
to take account of large fluctuations in the frustrated system. 
The method is a standard one in which 
configurations of classical localized spins are sampled by Monte Carlo procedure; 
the Monte Carlo weight is calculated by the exact diagonalization of 
the fermion Hamiltonian matrix for a given spin configuration. 
The bottleneck of the calculations is the exact diagonalization,
which usually limits the accessible system sizes to several hundreds sites.
An improved method based on the polynomial expansion of the density of states 
was developed by the authors
\cite{Motome1999,Furukawa2004}, 
but for frustrated systems, in general, the method becomes less efficient
because of singular form of the density of states: 
Indeed, in the present pyrochlore case, the density of states exhibits 
a $\delta$-functional peak due to two flat bands ($\omega/t = 2$) 
as well as two van-Hove singularities ($\omega/t = 0$ and $-4$) 
in the non-interacting model as shown in Fig.~\ref{fig1}(b).
Therefore we here employ the standard method.

To enable systematic analysis within 
the limited system sizes, 
we apply a technique of averaging over the twisted boundary conditions
\cite{Poilblanc1991,Gros1992}. 
In this technique, a twisted boundary condition is imposed 
with replacing the transfer integral $t$ by 
$t \exp(i \mbox{\boldmath $\phi$} \cdot \mbox{\boldmath $\delta$}_{ij})$,
where {\boldmath $\phi$} denotes a magnetic flux and 
$\mbox{\boldmath $\delta$}_{ij}$ represents the vector connecting the nearest-neighbor sites 
$\langle ij \rangle$. 
The average is taken by the integral over $\mbox{\boldmath $\phi$}$,
which is approximately calculated by the sum over $N_\phi$ grid points. 
It has been shown that the procedure reduces finite size effects 
originating from the discreteness of the wave numbers. 

It is shown for the model (\ref{eq:H}) that 
in the limit of $N_\phi \to \infty$, the averaging procedure 
for a state of a finite size cluster with a given set of spin configurations
provides results for the infinite size system which consists of a periodic array of the finite size cluster. 
For example, for the perfectly ordered ferromagnetic state, 
the averaging procedure with $N_\phi \to \infty$ 
for any finite size cluster gives the exact result 
in the thermodynamic limit. 
In order to take this advantage, we apply the averaging procedure 
to each Monte Carlo snapshot for calculating the Monte Carlo weight. 
This corresponds to ensemble average over the independent systems 
with different boundary conditions. 
In the following calculations, we take the average over the grid points 
with $\mbox{\boldmath $\phi$} = ( (2m_x - 1) \Delta\phi_x, 
(2m_y - 1) \Delta\phi_y, (2m_z - 1) \Delta\phi_z)$, 
where $\Delta\phi_\nu = \pi/2L_\nu l_\nu$ and 
$m_\nu = 1,2, \cdots, l_\nu$ ($\nu = x,y,z$).
Here, $N_\phi = l_x \! \times \! l_y \! \times \! l_z$ and
$L_\nu$ is a linear dimension of the system measured in the cubic unit cell, 
i.e., the total number of sites $N_{\rm s} = L_x \! \times \! L_y \! \times \! L_z \! \times \! 16$. 

\begin{figure}[h]
\begin{minipage}{37pc}
\includegraphics[width=28pc]{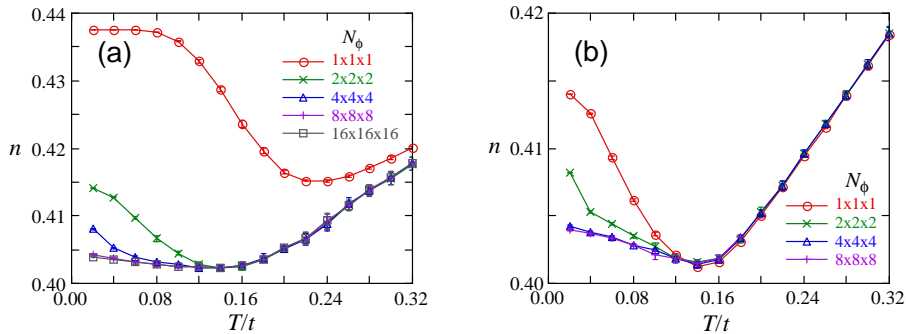}
\caption{
$T$ dependence of the electron density $n$ at $\mu=0$ 
for different numbers of grid points $N_\phi$. 
(a) The system size $N_{\rm s} = 1 \! \times 1 \! \times \! 1 \! \times \! 16$
and (b) $N_{\rm s} = 2 \! \times 2 \! \times \! 2 \! \times \! 16$.
The lines are guides for the eyes.
}
\label{fig2}
\end{minipage}
\end{figure}

\begin{figure}[h]
\begin{minipage}{37pc}
\includegraphics[width=28pc]{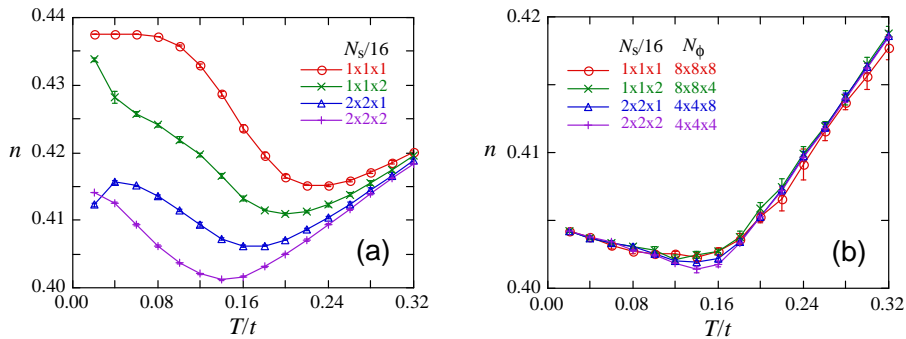}
\caption{
$T$ dependence of $n$ at $\mu=0$ for different system sizes $N_{\rm s}$
(a) without taking average 
($N_\phi = 1 \! \times 1 \! \times \! 1$)
and (b) with taking average $N_\phi$ 
(the numbers are shown in the legend). 
The lines are guides for the eyes.
}
\label{fig3}
\end{minipage}
\end{figure}

We demonstrate here the efficiency of 
the averaging technique by calculating the electron density $n$ as a function of $T$. 
We show the Monte Carlo results at $\mu=0$  in Figs.~\ref{fig2} and \ref{fig3} as an example. 
In Fig.~\ref{fig2}, we present $N_\phi$ dependence for two different system sizes.
In both cases, the results converge onto a single curve as increasing $N_\phi$. 
Furthermore, the necessary $N_\phi$ for the convergence 
becomes smaller as the system size $N_{\rm s}$ increases. 
Figure~\ref{fig3} shows $N_{\rm s}$ dependence with and without taking averages. 
In the case without taking averages, 
the results are largely scattered for different 
system sizes as shown in Fig.~\ref{fig3}(a). 
In contrast, as in Fig.~\ref{fig3}(b),
when we take averages over a sufficient number of $N_\phi$
chosen for each system size, all the results converge onto a single curve 
and the finite size effects are sufficiently small,  
even for the rectangular-shaped clusters. 
We have checked the efficiency in wide parameter regions of $\mu$ and $T$ 
and confirmed that 
the averaging procedure is efficient enough to suppress the finite size effects
coming from the discreteness of the wave numbers. 
We adopt $N_\phi$ used in Fig.~\ref{fig3}(b) for each $N_{\rm s}$ in the following calculations.

\vspace{-4mm}

\section{Results and Discussion}
\begin{figure}[h]
\begin{minipage}{37pc}
\includegraphics[width=32pc]{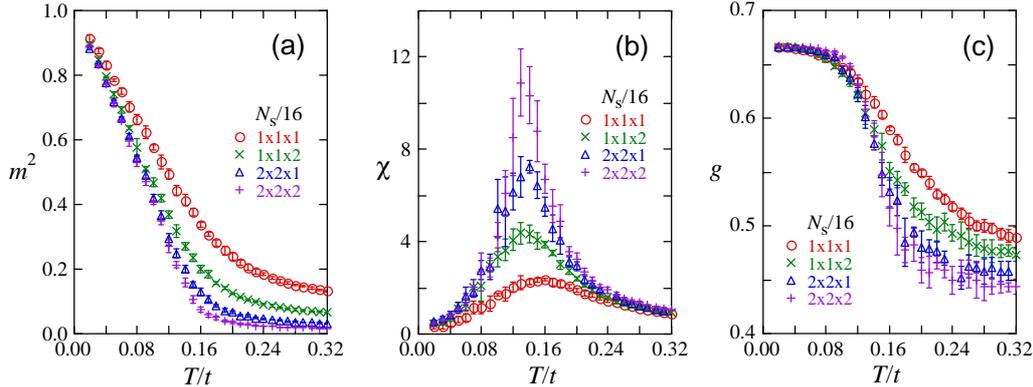}
\caption{
$T$ dependences of (a) square of total magnetization per site,
(b) uniform magnetic susceptibility, and
(c) Binder parameter for the total magnetization per site.
The electron density is fixed at $n=0.5$ by controlling $\mu$.
}
\label{fig4}
\end{minipage}
\end{figure}

Applying the method above, we investigate the magnetic behavior of the model (\ref{eq:H}) 
in the limit of $J_{\rm H}/t \to \infty$ at the electron density $n=0.5$. 
The results are presented in Fig.~\ref{fig4}. 
At $T/t \simeq 0.14$, the square of total magnetization per site 
${\mathbf m} = \sum_i {\mathbf S}_i / N_{\rm s}$ grows rapidly 
[Fig.~\ref{fig4}(a)] and 
the uniform magnetic susceptibility 
$\chi = ( \langle | {\mathbf m} |^2 \rangle - \langle | {\mathbf m} | \rangle^2 ) N_{\rm s} / T$ 
exhibits a peak [Fig.~\ref{fig4}(b)], signaling a ferromagnetic transition. 
We estimate $T_{\rm c}$ from the crossing point of the Binder parameter 
$g = 1 - \langle ( | {\mathbf m} |^2 )^2 \rangle / 3 \langle | {\mathbf m} |^2 \rangle^2$ 
for different system sizes: 
The estimate is $T_{\rm c}/t = 0.135(15)$. 

The value of $T_{\rm c}$ is quite similar to that for the cubic lattice model, 
$T_{\rm c}/t =  0.136(2)$
\cite{Motome2003b}. 
This is reasonable because the DE ferromagnetism 
is governed by 
the kinetic energy of electrons, and 
does not strongly depend on the details of the lattice structure. 

Our results demonstrate that systematic and quantitative study of the phase diagram is feasible 
within relatively small size clusters 
by implementing the averaging procedure over twisted boundary conditions 
in the Monte Carlo calculations. 
This gives a starting point for further study of 
frustrated DE-based models
and for understanding the physics of pyrochlore-based ferromagnets 
such as Mo pyrochlore oxides. 
Studies of phase competitions in the model including the super-exchange coupling 
between localized moments are in progress. 

This work was supported by Grant-in-Aid for Scientific Research on Priority Areas 
(Nos. 17071003, 19052008), Global COE Program ``the Physical Sciences Frontier" and
by the Next Generation Super Computing Project, Nanoscience Program, MEXT, Japan.

\end{document}